# CONTENT ANALYSIS APPLICATION IN NURSING: A SYNTHETIC KNOWLEDGE SYNTHESIS META-STUDY

Helena Blažun Vošner[1,2,3], Peter Kokol[4], Jernej Završnik[1,3,5,6] Danica Železnik[2]

[1] Zdravstveni dom dr. Adolfa Drolca Maribor, Ulica talcev 9, 2000 Maribor
[2] Fakulteta za zdravstvene in socialne vede Slovenj Gradec, Glavni trg 1, 2380 Slovenj Gradec
[3] Alma Mater Europaea, Slovenska ulica 17, 2000 Maribor
[4] Univerza v Mariboru, Fakulteta za elektrotehniko, računalništvo in informatiko, koroška cesta 46, 2000 Maribor
[5] Univerza v Mariboru, Fakulteta za naravoslovje in matematiko, Koroška cesta 160, 2000 Maribor
[6] Znanstveno raziskovalno središče Koper, Garibaldijeva ulica 1, 6000 Koper

***Abstract:***
***Theoretical issues:*** *With the explosive growth in the research literature production, the need for new approaches to structure knowledge emerged.*
***Method:*** *Synthetic content analysis was used in our meta-study.*
***Results and discussion:*** *Our meta-study showed that content analysis is frequently used in nursing research in a very wide spectrum of applications. The trend of its use is positive and it is used globally in a variety of research settings. The synthetic content analysis used in our study showed to be a very helpful tool in performing knowledge synthesis, replacing many of the routine activities of conventional synthesis with automated activities this making such studies more economically viable and easier to perform.*

## 1. Introduction

With the explosive growth in the research literature production, the need for new approaches to structure knowledge emerged (Merton, 1985). Capability to analyse vast amounts of publications on macroscopic and microscopic levels, together with domain independence classifies bibliometrics as one of such approaches (Tejasen, 2016). Bibliometric was defined by Pritchard (1969), as "the application of mathematical and statistical methods to books and other media of communication" and by Hawkins (2001) as "the quantitative analysis of the bibliographic features of a body of literature".Bibliometrics has become increasingly popular (Pesta, Fuerst & Kirkegaard, 2018), especially in medical and nursing research (Kokol & Blažun Vošner, 2019; Kokol, Vošner & Završnik). Among almost 20000 bibliometric papers indexed in Scopus more than one third are from the medical and nursing fields.

Another approach to deal with the explosive growth in the research literature production is knowledge synthesis. Knowledge synthesis roots dates back more than 120 years, however become more popular in the 1960s (Chalmers, Hedges & Cooper, 2002), and even more commonly used toward the end of the millennia with the introduction of the evidence-based practice (Tricco, Tetzlaff & Moher, 2011; Whittemore et al., 2014). One of the more popular knowledge synthesis methods especially in nursing is content analysis. Its main advantages are that it is content-sensitive, highly flexible, and can be used to analyse many types of data, either in inductive or deductive manner (Kyngäs, Mikkonen & Kääriäinen, 2020).

Triangulating bibliometrics and knowledge synthesis leads to the so called synthetic knowledge synthesis (Blažun Vošner et al., 2017). The objective of our paper is present the outputs of a meta-study, in which we used synthetic content analysis to analyse the scope of application of content analysis in nursing. Thus, our aim was to identify the descriptive and spatial bibliometric patterns like trends, most prolific countries, institutions, source titles and themes.

## 2. Methodology

Synthetic content analysis was performed using the algorithm below:

1. Harvest the research publications concerning medical software quality to represent the content to analyse.
2. Condense and code the content using text mining. Authors keywords were selected as candidate codes, cause they most concisely present the content of a publication (Železnik, Blažun Vošner & Kokol, 2017))
3. Map the codes using bibliometric mapping and induce a clustered author keywords landscape.
4. Analyse the connections between the codes in individual clusters and map them into sub-categories.
5. Identify interesting issues and the scope of categories (type of care, patient or illness)
6. Analyse sub-categories and label clusters with themes

The search was performed in Web of Science Core Collection on January 17th, 2021, for the period 2011-2020. We used the search string "Content analysis" limited to the Research Area = Nursing. The bibliometric mapping and text mining were performed using the VOSViewer software 1.6.15 (van Eck & Waltman, 2014).

## 3. Results and discussion

Number of the harvested publications was 5119. Among them there were 4903 original articles, 180 reviews, 26 conference papers, nine editorials and one correction. Figure 1. Shows that during last ten years the number of papers steadily increased from 300 to 725, reaching the peak in 2019 with 805 papers.

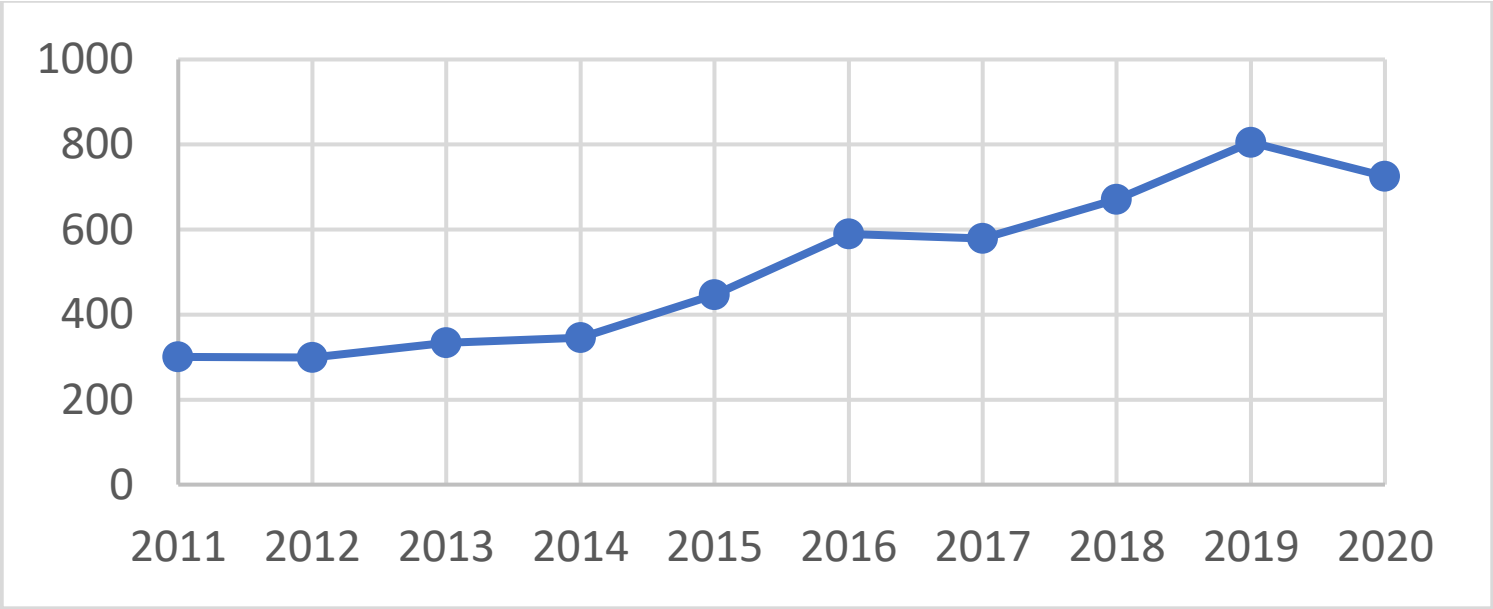

Figure 1. The dynamic of literature production

### *3.1.Spatial bibliometrics*

Most productive countries, organisations and source titles are presented in Tabel 1 to 3. As in most of other scientific disciplines the USA is the most productive country, and as usual in nursing, England and Scandinavian countries are among the top 10 most productive countries. Similarly, the organisation from those countries are also among the most productive ones. The papers are published in respected international journals mostly with high impact factors.

Table 1: 10 most productive countries

| County | Number of publications |
|---|---|
| USA | 1165 |
| SWEDEN | 975 |
| BRAZIL | 692 |
| AUSTRALIA | 393 |
| IRAN | 323 |
| FINLAND | 268 |
| NORWAY | 261 |
| CANADA | 255 |
| ENGLAND | 214 |
| PEOPLES R CHINA | 139 |

Table 2: 10 most productive organisations

| Organizations | Number of publications |
|---|---|
| KAROLINSKA INSTITUTET | 229 |
| UNIVERSITY OF GOTHENBURG | 152 |
| UNIVERSIDADE DE SAO PAULO | 138 |
| LUND UNIVERSITY | 118 |
| UMEA UNIVERSITY | 115 |
| UPPSALA UNIVERSITY | 110 |
| TEHRAN UNIVERSITY OF MEDICAL SCIENCES | 91 |
| UNIVERSITY OF TURKU | 88 |
| LINKOPING UNIVERSITY | 86 |
| UNIVERSITY OF EASTERN FINLAND | 76 |

Table 3: 10 most prolific journals

| Source Titles | Number of publications |
|---|---|
| JOURNAL OF CLINICAL NURSING | 376 |
| SCANDINAVIAN JOURNAL OF CARING SCIENCES | 221 |
| REVISTA BRASILEIRA DE ENFERMAGEM | 168 |
| REVISTA DE PESQUISA CUIDADO E FUNDAMENTAL ONLINE | 156 |
| NURSE EDUCATION TODAY | 155 |
| JOURNAL OF ADVANCED NURSING | 152 |
| REVISTA DA ESCOLA DE ENFERMAGEM DA USP | 127 |
| MIDWIFERY | 115 |
| NURSING ETHICS | 111 |
| JOURNAL OF NURSING MANAGEMENT | 100 |

### *3.2.Synthetic content analysis*

The results of the synthetic content analysis is shown Figure 2 and Table 4. Five themes emerged, ranging from long term and acute care, use of technology in nursing, education and quality of life. The content analysis was used for analysis of nursing care in wide spectrum of illness, from new-borns to elderly patients, almost all types of care and various health institution settings. It was mostly used in inductive manner. The content analysis was also as part of different other knowledge synthesis methods, for example Literature reviews (n=305), Systematic reviews (n=56), Integrative reviews (n=43) and Scoping reviews (n=10).

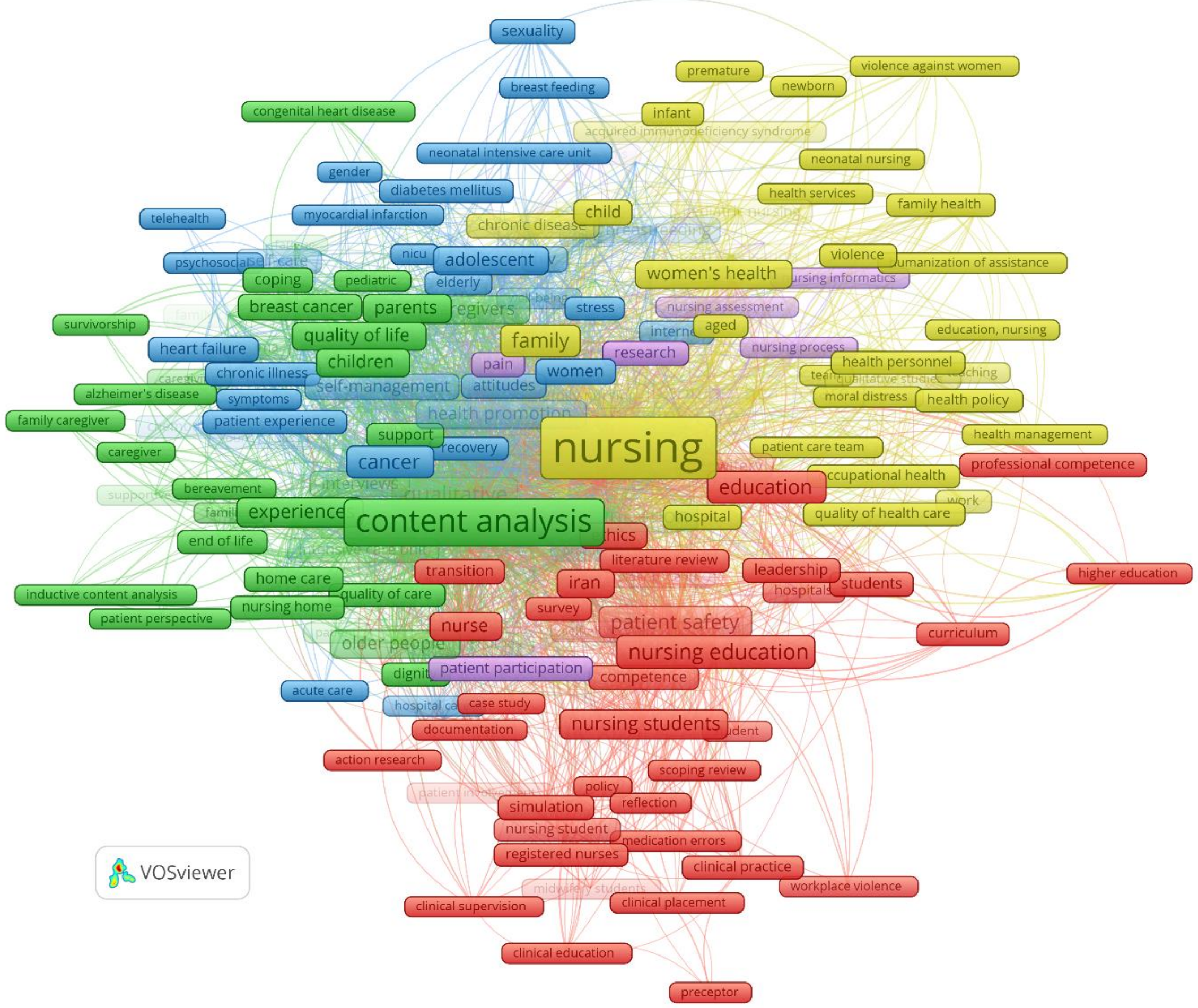

Figure 2. Author keywords cluster landscape

Table 4: Content analysis of “content analysis nursing publications” based on the author keyword cluster landscape (Figure 1)

| **Theme** | **Colour** | **More frequent codes**<br>**Interesting issues** | **Categories** | **Patients/care/illness** |
|---|---|---|---|---|
| Nursing education to improve nursing care | Red | Nurses (231), Nursing education (113), Patient safety (102), Midwifery (49), Family centred care (15), primary care (34)<br>Perception, perceptron. Medication errors | Clinical and theoretical nursing education, Nursing education using simulations, improving patient safety with teamwork and reducing medication errors, qualitative research in professional competences for nursing leadership, Improved competency with midwifery education, perception of health care stake holders | Mothers, older patients, family centred care, maternity care, primary care, prenatal care, emergency care |
| Quality of life of elderly | Green | Content analysis (327), communications (119), quality of life (66), older people (66), children (68), dementia (56), caregivers (59). Nursing homes (45)<br>Dignity, safety | Communications in end of life care in nursing homes, coping with fatigue, caregivers improving quality of life using cultural competences, dignity in end of life care, safety of children, safety in critical care | Older people, children, parents, families, woman, young and old adults, cancer, intensive care, Alzheimer’s disease, HIV, dementia, Palliative care, end of life care, supportive care, home care, long term care |
| Long term care | Blue | Qualitative research (473), adolescent (110), cancer (104), breastfeeding (54), health promotion (59), self management (49)<br>Spirituality, religion, well-being, aging, internet, stress, sexuality | Empowerment of adolescents suffering heart diseases and diabetes with self care/management, breastfeeding in neonatal intensive care unit, well being in context of gender, caner and spirituality in cancer patients, stress and internet, health promotion for person centerd care | Adolescent, elderly, older adults, fathers, mothers, Hospital care, acute care, neonatal intensive care, heart failure, diabetes, chronic illness |
| Relations between speciality nursing discipline in holistic family care | Yellow | Nursing (650), nursing care (138), family (96), mental health (64), primary health care (97), woman health (64), child (50), health education (46), pediatric nursing (32), chronic diseases (32)<br>Quality of health care, moral distress, violence, adaption, humanisation of assistance | Comprehensive health care for family, woman and child, complexity of family health nursing, issues of primary health care in family nursing, violence against woman in woman health, pediatric nursing in children chronic diseases, importance of health education in family health, moral distress in nursing decision making during burnout, health policies in humanisation of assistance | Child, family, infant, premature Nursing care, mental health, chronic diseases, family nursing, occupational health, intensive care, psychiatric nursing, public health nursing, emergency nursing |
| Technology in nursing care | Violet | Nursing research (50), pain (41), evidence based practice (31), Stroke (31), Nursing intervention (35), patient participation (32) prevention (26), rehabilitation (24)<br>Nursing informatics, Technology, Social media, Covid 19 | Social media in nursing research, nursing interventions for pain management, rehabilitation of stroke patients, patient participation in clinical decision making, prevention for sexual health, nursing informatics in nursing practice | Stroke, Mental health nursing |

## 4. Conclusion

Our meta-study showed that content analysis is frequently used in nursing research in a very wide spectrum of applications. The trend of its is positive and content analysis is used globally in a variety of research setting. The synthetic content analysis used in our study showed to be a very helpful tool in performing knowledge synthesis, replacing many of the routine activities of conventional synthesis with automated activities this making such studies more economically viable and easier to perform.